\documentclass[aps,prl,reprint,groupedaddress]{revtex4-1}
\usepackage{url}
\usepackage[dvips]{graphicx}
\usepackage{amsthm}
\usepackage{amsmath}
\usepackage{physics}
\usepackage{here}
\usepackage{amssymb}
\usepackage{dcolumn}
\usepackage{bm}
\usepackage{subfigure}
\usepackage{bbold}

\newtheorem{condition}{Condition}

\allowdisplaybreaks[0]

\begin{document}
\title{Fundamental bound on the power of quantum machines}

\author{Kosuke Ito}
\affiliation{Graduate School of Mathematics, Nagoya University, Furocho, Chikusa-ku, Nagoya 464-8602, Japan}
\email{m13007a@math.nagoya-u.ac.jp}
\author{Takayuki Miyadera}
\affiliation{Department of Nuclear Engineering, Kyoto University, Kyoto daigaku-katsura, Nishikyo-ku, Kyoto 615-8530, Japan}
\email{miyadera@nucleng.kyoto-u.ac.jp}
\begin{abstract}
 Giving a universal upper bound on the power output of heat engines is a long-standing open problem.
 We tackle this problem for generic quantum machines in a self-contained formulation by carefully including the switching process of the interaction.
 In this way, we show a fundamental upper bound on the power associated with the energy-time uncertainty principle.
 As a result, the energy fluctuation of the controller is identified as a necessary resource for producing the power.
 This bound implies a trade-off between the power and `noise' in the energy, which yields an estimation on the time scale for detectable work extraction.
 Ideal clock-driven model of autonomous quantum machine gives a concrete demonstration of our bound.
\end{abstract}

\maketitle

 The recent surge in investigations of thermodynamics of quantum systems has revealed universal bounds on the extractable work via quantum heat engines in various ways \cite{sai-qthermo2015,Gelbwaser-Klimovsky:2015ab,1367-2630-18-1-011002}.
 However, little is known about the universal characterization of the time it takes, namely the power, work per unit time, of heat engines.
 No matter how large the extracted work is, it makes no sense in practice if it takes forever,
 as the Carnot engine is practically useless because of its vanishing power.
 In fact, intensive research has been done on the finite-time thermodynamics
 in relation with the efficiency \cite{Chambadal:1957aa,Novikov1957,doi:10.1119/1.10023,PhysRevLett.105.150603,PhysRevLett.117.190601,PhysRevE.96.022138,PhysRevLett.106.230602}.
 Especially, an explicit trade-off relation between the power and the efficiency was recently derived for classical \cite{PhysRevLett.117.190601} and quantum \cite{PhysRevE.96.022138} Markovian heat engines,
 which showed that Carnot efficiency can never be achieved at finite power with Markovianity
 \footnote{Incompatibility between finite power and Carnot efficiency was also proved for another class of quantum heat engines where Lieb-Robinson bound is applicable \cite{PhysRevE.96.022138}}.
 Despite these enormous progress, universal understanding of bounds on the power is still missing.
 Indeed, there is even room for achieving Carnot efficiency at finite power at present \cite{PhysRevLett.106.230602}.
 Actually, it has been yet unclear in what level, what characterizes the limits on the power.

 Is any external resource required to produce the power?
 The majority of conventional studies on the finite-time quantum thermodynamics deal with externally controlled Hamiltonians \cite{PhysRevE.83.041117,e18060205,1367-2630-18-7-075019,PhysRevLett.118.100602}.
 In such approaches,
 it is unclear whether some external resource implicitly contributes to producing the power because they are not self-contained.
 As an alternative approach, one may consider an operation on the system and an explicitly included work storage system via a time independent interaction,
 as with some approaches in quantum thermodynamics \cite{Horodecki:2013aa,Skrzypczyk:2014aa,Aberg:2014aa,PhysRevA.95.032132}.
 Then, let us do so firstly.
 
 We consider two harmonic oscillators as the system and the work storage whose respective Hamiltonians are
 $H_S=\hbar\omega a^{\dagger}a$ and $H_W=\hbar\omega b^{\dagger}b$
 , where $a$ and $b$ are the respective annihilation operators of the system and the work storage.
 They are prepared in a product state $\rho\otimes \sigma$ at time $t=0$, and
 the energy is extracted from the system oscillator to the work storage through the interaction Hamiltonian $V= g (a^{\dagger}b + a b^{\dagger})$ until $t=\tau$ with the coupling constant $g$.
 When the initial state $\sigma$ of the work storage is diagonal in the energy eigenstates,
 the amount of the extracted average energy is calculated as $(\langle H_S \rangle - \langle H_W\rangle)(1-\cos(2g\tau/\hbar))/2 $, where $\langle H_S \rangle $ and $ \langle H_W\rangle$ are the initial average energies of the system and the work storage, respectively.
 As long as $\langle H_S \rangle > \langle H_W\rangle$, the positive work $\langle H_S \rangle - \langle H_W\rangle$ is obtained at $\tau = \pi\hbar/(2g)$, which can be arbitrarily short by taking large $g$.
 Therefore, it turns out that any large power can be obtained with any diagonal state of the work storage even including the ground state.
 Then, does this mean that any external resource is unnecessary?
 The answer is {\it no}, because this process is still {\it not} truly self-contained in the following sense.
 Carefully looking at this model, one may perceive that the switching on and off of the interaction are externally given at $t=0$ and $t=\tau$ respectively, so that the interaction Hamiltonian is actually time dependent as $V(t)= 0 \; (t<0,t>\tau),\; V(t)= V \;(0\leq t \leq \tau)$.

 In this letter, we show a fundamental upper bound on the power produced by a quantum machine in a self-contained formulation.
 As shown below, it is essential for characterizing the bound to include on-and-off switching process.
 The bound reveals that the energy fluctuation of the external controller is a necessary resource for producing the power.
 
\paragraph{Autonomous quantum machine.---}
We deal with generic quantum machines to exchange the energy between a quantum system $\mathcal{H}_S$ and an agent $\mathcal{H}_A$ whose respective Hamiltonians are $H_S$ and $H_A$.
Especially, we consider a self-contained formulation of the quantum machine with the time-independent total Hamiltonian $H = H_0 + V := H_S\otimes \mathbb{1}_S + \mathbb{1}_A\otimes H_A + V$ which describes an autonomous interaction.
That is, we assume that the initial state $\sigma_A $ of the agent and the Hamiltonian $H$ satisfy the following condition:
 \begin{condition}[No interaction up to time $t_0$ \cite{Miyadera2016}]\label{cond1}
  For any state $\rho$ of the system $\mathcal{H}_S$ and any time $t\leq t_0$ before the initial time $t_0$,
  the commutativity
$
 [V,\rho\otimes\sigma_A(t)]=0
  $
 is satisfied,
 where $\sigma_A(t):= e^{-i\frac{H_A (t-t_0)}{\hbar}}\sigma_A e^{i\frac{H_A (t-t_0)}{\hbar}}$.
\end{condition}
This condition guarantees that
the two systems $\mathcal{H}_S$ and $\mathcal{H}_A$ are separated from each other until the switching-on time after $t_0$.
For simplicity, we set $t_0=0$ from now on.
It was shown \cite{Miyadera2016} that Condition \ref{cond1} is equivalent to that
\begin{align}
  e^{-i\frac{H t}{\hbar}} (\rho\otimes \sigma_A) e^{i\frac{H t}{\hbar}}
 =
 e^{-i\frac{H_S t}{\hbar}}\rho e^{i\frac{H_S t}{\hbar}} \otimes \sigma_A(t)
\end{align}
holds for any state $\rho$ of $\mathcal{H}_S$ and $t\leq 0$.
Then, the system and the agent describe a generic autonomous quantum machine where the interaction is automatically switched-on after $t=0$.
In this way, all the components involved in the process including the ``switch'' of the interaction are contained in our formulation.

 We denote the initial state of the system by $\rho_S$, and the time evolution $e^{-i\frac{H t}{\hbar}} (\rho_S \otimes \sigma_A) e^{i\frac{H t}{\hbar}}$ of the total system by $\Theta(t)$.
 We focus on the mean work $W:=\tr H_S (\rho_S - \rho_S'(\tau))$, where $\rho_S'(\tau):= \tr_A \Theta(\tau)$ is the final state of the system at the final time $\tau > 0$.
Then, the mean power $P$ is defined as $P:= W/\tau$.
 To interpret $W$ as the extracted work from the system to the agent, we assume the average energy conservation
 \begin{align}
  \tr (H_S+H_A)\Theta(\tau)=\tr (H_S+H_A)\Theta(0) \label{av_cons}
 \end{align}
 together with the switch-off condition:
 \begin{align}
  [V,\Theta(t)]=0 \quad(t\geq \tau).\label{off_cond}
 \end{align}
 Condition \eqref{av_cons} guarantees that
the interaction Hamiltonian $V$ just redistributes the energy between the system and the agent, so that
the total energy $H_S+H_A$ is not affected
\footnote{We cannot simply impose the energy conservation $[H_S+H_A,V]=0$ because the dynamics becomes trivial in combination with Condition \ref{cond1}.
In fact, the energy conservation implies
$e^{-i\frac{(H_S+H_A+V) t}{\hbar}}= e^{-i\frac{(H_S+H_A) t}{\hbar}} e^{-i\frac{V t}{\hbar}}$,
so that
$
\Theta (t)
 = e^{-i\frac{(H_S+H_A) t}{\hbar}} e^{-i\frac{V t}{\hbar}}
 (\rho_S\otimes \sigma_A)
 e^{i\frac{V t}{\hbar}} e^{i\frac{(H_S+H_A) t}{\hbar}} 
 = e^{-i\frac{(H_S+H_A) t}{\hbar}}
 (\rho_S\otimes \sigma_A)
 e^{i\frac{(H_S+H_A) t}{\hbar}}
 $
 follows from $[V,\rho_S\otimes \sigma_A]=0$.}.
 Otherwise, the system and agent become no longer self-contained because of some external degree of freedom associated with the change in the interaction energy.
 Condition \eqref{off_cond} ensures that the interaction is turned off after the interaction time $\tau$ so that the energies $H_A$ and $H_S$ will respectively remain unchanged after that time.
 In particular, if $V\Theta(t)=0$ holds for $t\leq 0$ and $t\geq \tau$, Condition \ref{cond1} and (\ref{av_cons}), (\ref{off_cond}) are all satisfied.

\paragraph{Fundamental power bound.---}
Our first main result is the following fundamental bound on the power of quantum machines:
 \begin{align}
  |P| \leq \frac{2\|H_S\| \Delta H_A }{\hbar}\label{pb_f}
 \end{align}
 in terms of the energy fluctuation $\Delta H_A := \sqrt{\tr H_A^2 \sigma_A - (\tr H_A \sigma_A)^2}$ of the agent, where $\|B\|$ is the operator norm of an operator $B$
 \footnote{Note that our bound \eqref{pb_f} holds under Condition \ref{cond1} alone without assuming the energy conservation \eqref{av_cons} and the switch-off condition \eqref{off_cond}.
  They are only necessary to interpret $W$ as the exchanged work between the system and the agent. We also remark that this bound gives limits on both work extracted from and done on the system as $-2 \norm{H_{S}} \Delta H_A/\hbar \leq P \leq 2 \norm{H_{S}} \Delta H_A/\hbar$.}.
 This bound identifies the energy fluctuation of the external controller as a necessary resource for producing the power.
 We give a detailed proof of \eqref{pb_f} later.
    
 Our bound (\ref{pb_f}) is a kind of quantum speed limit (QSL) on the power associated with the energy-time uncertainty principle \cite{1751-8121-50-45-453001}.
 Actually, our bound is derived from the following stronger bound in terms of the trace norm
 $\|[H_A,\sigma_A ]\|_1$ of the commutator $[H_A,\sigma_A]$:
   \begin{align}
  |P| \leq \frac{\|H_S\|\|[H_A,\sigma_A ]\|_1}{\hbar},\label{pb_1}
   \end{align}
 which essentially follows from a recently derived QSL by Marvian \cite{PhysRevA.93.052331}.
 This bound \eqref{pb_1} implies that not only the fluctuation but also sufficient amount of quantum superposition in the energy eigenstates is necessary to extract the power.
 In fact, even though $\Delta H_A$ is large due to only a classical mixture of energy eigenstates $\ket{E_{A,i}}$ as $\sigma_A=\sum_i p_i\ketbra{E_{A,i}}$,
 the possible power is zero according to \eqref{pb_1} since $[H_A,\sigma_A]=0$.

 \begin{figure}[!t]
\centering
\includegraphics[clip ,width=3.3in]{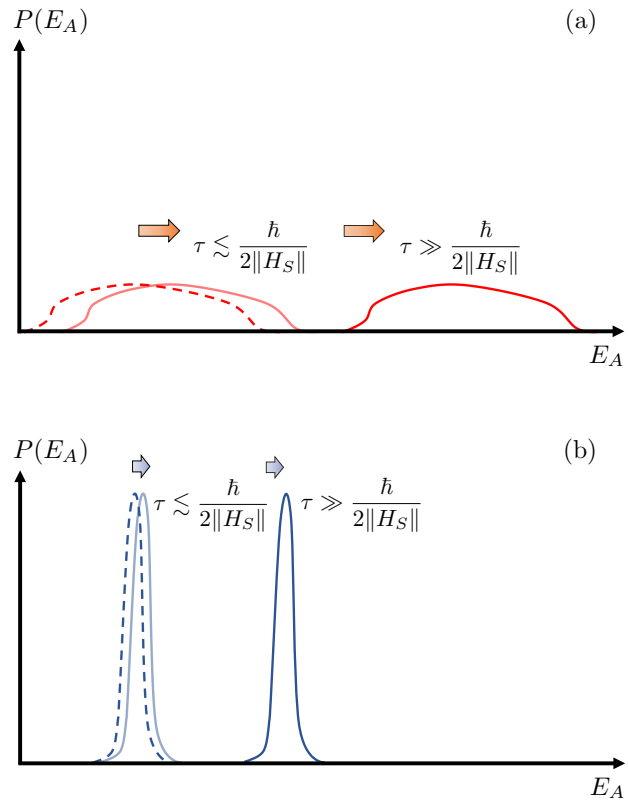}
 \caption{A sketch of the time evolution of the probability distribution $P(E_A)$ of outcomes $E_A$ of the energy measurement of the agent when the work fluctuation is small enough.
 Unless $W\gg \Delta H_A$ (i.e.~for time scale $\tau \lesssim \hbar/(2\|H_S\|$),
 the work is hard to detect due to the great deal of overlap between the initial (dashed curves in both graphs) and the final distribution (thin colored (gray) curves in both graphs).
 In the case (a) where the fluctuation is large, large amount of work is required to apart from the initial distribution though the speed can be fast. (b) When the fluctuation is small, the speed is slow although small work is sufficient.
 In consequence, long enough time $\tau \gg \hbar/(2\|H_S\|)$ is required regardless of the fluctuation.}
\label{figure1}
\end{figure}
 
 Especially to focus on the work extraction from the system to the agent where the agent is the work storage, our bound \eqref{pb_f} implies the trade-off between the power and `noise'.
That is, as large power is produced, the resulting signal of the work tends to be hidden behind the inevitable large `noise' of the fluctuation implied by bound \eqref{pb_f}.
Such an argument is known in the context of Einstein photon box \cite{Busch1990,Busch:2008aa}.
Recently, similar trade-off relations were also discussed on relations between the precision of a unitary operation and detectability of the output work \cite{PhysRevA.95.032132,1709.06920}.
As a first application of bound \eqref{pb_f} from the perspective of such a trade-off, let us consider the time scale required to extract the work
under the assumption of small enough work fluctuation.
In this case, the energy distribution of the agent moves almost parallelly as in Fig.~\ref{figure1}.
To detect the work in this case, the change $W$ in the energy have to be larger enough than the energy fluctuation $\Delta H_A$ as
$\Delta H_A \ll W$.
This fact and bound \eqref{pb_f} yield the necessary condition $\Delta H_A \ll 2\|H_S\| \Delta H_A \tau / \hbar$ for detectable work extraction.
Hence, we obtain the estimation
\begin{align}
 \tau \gg \frac{\hbar}{2\|H_S\|}\label{TS}
\end{align}
of its time scale.
The trade-off results in this inevitable speed limit regardless of the energy fluctuation of the agent but rather characterized by the system energy scale $\|H_S\|$ alone (Fig.~\ref{figure1}).

 Now, we prove inequality \eqref{pb_f}. We compare the time evolution
 of the initial state $\rho_S \otimes \sigma_A$
 with
 that of
 $\tau$-delayed agent
 $\rho_S \otimes \sigma_A (-\tau)$.
 With the same dynamics governed by $H$,
 the former yields $\Theta(\tau) = e^{-i\frac{H \tau}{\hbar}} (\rho_S \otimes \sigma_A) e^{i\frac{H \tau}{\hbar}}$,
 and the latter does $\rho_S(\tau)\otimes \sigma_A$ after the duration $\tau$, where $\rho_S(\tau):= e^{-i\frac{H_S \tau}{\hbar}}\rho_S e^{i\frac{H_S \tau}{\hbar}}$.
 From the unitary invariance of the trace norm, we have
$
  \|\Theta (\tau) - \rho_S (\tau)\otimes \sigma_A\|_1
  =
  \|\rho_S\otimes \sigma_A - \rho_S \otimes \sigma_A (-\tau)\|_1
  =
  \|\sigma_A - \sigma_A (-\tau)\|_1.
$
 In combination with this equation, the monotonicity
 $\|\Theta (\tau) - \rho_S (\tau)\otimes \sigma_A\|_1 \geq \|\rho_S'(\tau) - \rho_S (\tau)\|_1$
 of the trace distance with respect to the partial trace yields
 \begin{align}
  \|\sigma_A - \sigma_A (-\tau)\|_1 \geq \|\rho_S'(\tau) - \rho_S (\tau)\|_1.
 \end{align}
 Then, a quantum speed limit given by Marvian {\it et al.~}[(4.1)]\cite{PhysRevA.93.052331}
 implies
 \begin{align}
  \tau \geq \hbar\frac{\|\sigma_A - \sigma_A (-\tau)\|_1 }{\|[H_A,\sigma_A]\|_1}
  \geq \hbar\frac{\|\rho_S'(\tau) - \rho_S (\tau)\|_1 }{\|[H_A,\sigma_A]\|_1}.\label{qsl_1}
 \end{align}
 Since $H_S$ is conserved under the isolated dynamics of $\mathcal{H}_S$,
 $W= \tr H_S (\rho_S - \rho_S'(\tau)) = \tr H_S (\rho_S(\tau) - \rho_S'(\tau))$ holds.
 Thus, the inequality 
 $
 |W|
  \leq
  \|H_S\| \|\rho_S(\tau) - \rho_S'(\tau)\|_1 \label{w_in}
 $
 follows from
 $|\tr AB|\leq \|A\|\|B\|_1$ for any two operators $A$ and $B$.
 Combining this inequality with \eqref{qsl_1}, we obtain bound \eqref{pb_1}.
 Finally, the relation
 $
 \|[H_A,\sigma]\|_1 \leq 2 \Delta H_A
 $
  \cite{PhysRevA.93.052331}
  yields bound \eqref{pb_f}.
  
\paragraph{Ideal clock-driven quantum machine.---}
 As a typical model of an autonomous quantum machine, we consider the clock-driven quantum machine given by Malabarba {\it et al.}~\cite{1367-2630-17-4-045027}.
 In this model, the Hamiltonian of the agent is given by the momentum operator $P$ as $H_A=\nu P$, where we take $\nu= 1 [\mathrm{m}/\mathrm{s}]$.
 The system Hamiltonian $H_S$ is arbitrary as long as it is bounded.
 The interaction $V$ is defined by
  \begin{align}
  V= \int V_S(x)\otimes \ketbra{x} dx,
  \end{align}
 where $\ket{x}$ is the eigenstate with the eigenvalue $x$ of the position operator.
 The support $\mathrm{supp} V_S$ of $V_S$ is contained inside an interval of size $L$, namely $[0,L]$.
 The support of the initial state $\sigma_A$ of the agent in position is also contained inside a finite interval $[-K,0]$ of size $K$.
 Originally in \cite{1367-2630-17-4-045027}, this model was invented to reveal that energy conserving unitary driving of the system can be implemented by the agent
 without any work cost.
 Thus, this model was investigated under the commutativity $[H_S,V]=0$.
 On the other hand, we are interested in the energy exchange between the system and the agent.
 Thus, we rather assume non-commutativity $[H_S,V]\neq 0$.
 This is an idealized model of the clock driven quantum machine in the sense that the energy spectrum of the agent is doubly infinite and the initial state is completely confined in a finite region.
 
 In this model, the time evolution of the agent $\sigma_A(t)$ by $H_A$ alone is just the uniform motion $e^{-i\frac{H_A t}{\hbar}} = e^{-i\frac{ P t}{\hbar}}$ in position.
 Then, $V(\rho_S\otimes\sigma_A(t))=0$ is satisfied for any $\rho_S$ and $t\leq 0$ since
 the supports of $\sigma_A$ and $V_S$ have no intersection.
 Thus, Condition \ref{cond1} is satisfied.
 The global time evolution $\mathcal U(t) = e^{-i\frac{H t}{\hbar}}$
 is calculated as
 \begin{align}
  &\mathcal{U}(t)\nonumber\\
  =& e^{-i\frac{H_S t}{\hbar}} \int \mathcal{T} e^{-\frac{i}{\hbar} \int_0^{t} e^{-i\frac{s}{\hbar}H_S}V_S(x+s)e^{i\frac{s}{\hbar}H_S} ds}\otimes\ketbra{x+t}{x} dx,
 \end{align}
 where $\mathcal{T}$ is the time-ordered product.
 For simplicity, we suppose that the respective initial states $\rho_S$ and $\sigma_A$ of the system and the agent are pure, namely
 $\rho_S=\ketbra{\phi}$ and $\sigma_A=\ketbra{\psi}$, where $\ket{\psi}= \int \psi(x)\ket{x} dx$ with $\mathrm{supp}\psi \subset [-K,0]$.
 Then, the time evolution after the interaction time $t\geq \tau:= K+L$ becomes
 \begin{align}
  &\ket{\Theta(t)}:=\mathcal{U}(t)\ket{\phi}\otimes\ket{\psi}\nonumber\\
  =&
  \int_{-K}^{0}dx\; e^{-i\frac{H_S t}{\hbar}} U(x)\ket{\phi}
  \otimes \psi(x) \ket{x+t},
 \end{align}
 where $U(x):=e^{i\frac{H_S x}{\hbar}} U e^{-i\frac{H_S x}{\hbar}}$ with $U:= \mathcal{T}e^{-\frac{i}{\hbar} \int_0^L e^{-i\frac{s}{\hbar}H_S}V_S(s)e^{i\frac{s}{\hbar}H_S} ds}$.
 Thus, conditions (\ref{av_cons}) and \eqref{off_cond} are satisfied because of $V\ket{\Theta(t)}=0$ $(t\geq \tau)$ and $V\ket{\Theta(0)}=0$.
 Their validity is straightforwardly checked also for mixed states.
 For generic initial states $\rho_S$ and $\sigma_A:=\sum_i p_i \ketbra{\psi_i}$ of the system and the agent respectively, the final reduced state of the system is calculated as
 $
  \rho_S'(\tau)
  = \sum_i p_i \int dx |\psi_i (x)|^2 e^{-i\frac{H_S \tau}{\hbar}}U(x) \rho_S U(x)^{\dagger}e^{i\frac{H_S \tau}{\hbar}}.
 $
 Thus, the final energy of the system is
 \begin{align}
  &\tr H_S \sum_i p_i\int dx |\psi_i (x)|^2 e^{-i\frac{H_S \tau}{\hbar}}U(x) \rho_S U(x)^{\dagger}e^{i\frac{H_S \tau}{\hbar}}\nonumber\\
  =& \sum_i p_i \int dx |\psi_i (x)|^2 \tr H_S U e^{-i\frac{H_S x}{\hbar}} \rho_S e^{i\frac{H_S x}{\hbar}} U^{\dagger}.\label{fin_energy}
 \end{align}
 Especially, if $\rho_S$ is block-diagonal in energy eigenspaces as $[\rho_S, H_S]=0$,
 the final energy \eqref{fin_energy} coincides with the energy $\tr H_S U \rho_S U^{\dagger}$ obtained by the unitary operation $U$ independently of the initial state of the agent.
 Thus, the work extraction by an arbitrary unitary $U$ is realized for such an initial state
 by appropriately choosing $V_S$.
 For example, we can chose $V_S$ as
 \begin{align}
  V_S(s) = i\hbar f(s) e^{i\frac{s}{\hbar}H_S}(\log U) e^{-i\frac{s}{\hbar}H_S},\label{VS}
 \end{align}
 where $f$ is an arbitrary real function satisfying $\mathrm{supp}f\subset [0,L]$ and $\int f(s) ds = 1$.

 Now, let us consider how large power is attained in relation with our bound \eqref{pb_f} in this model.
 At first, we focus on how small fluctuation of the agent can be realized under the fixed size $L$ of the support of the initial state.
 The variance $\Delta H_A^2$ is concave as a function of density matrices of the agent.
 Therefore, it is sufficient to minimize the energy variance among the pure states since any density matrix can be decomposed into a convex combination of pure states.
 Then, we find an optimal initial pure state $\sigma_A=\ketbra{\psi}$
 which minimizes the variance
 $\Delta H_A^2 = \bra{\psi}H_A^2\ket{\psi} - \bra{\psi}H_A\ket{\psi}^2
 =\hbar^2 (\int |\psi'(x)|^2 dx + (\int \psi(x)^*\psi'(x) dx)^2)$.
 By the polar form $\psi(x)=r(x)e^{i\theta(x)}$ of the wave function, the variance may be written as $\hbar^2 (\int r'(x)^2 dx + \int r(x)^2\theta(x)^2 dx - (\int r(x)^2 \theta(x) dx)^2)$.
 Since the latter two terms $\int r(x)^2\theta(x)^2 dx - (\int r(x)^2 \theta(x) dx)^2$ is the variance of $\theta(x)$ under the probability density $r(x)^2$,
 it is enough to minimize $\int r'(x)^2 dx$ and take $\theta(x)\equiv 0$.
 This is done by solving the variational problem of the functional $\int_{-L}^0 r'(x)^2 dx$ of $r(x)$ on $[-L,0]$ under the constraint $\int r(x)^2 dx =1$ and the boundary condition $r(-L)=r(0)=0$.
 As a result, we obtain the optimal wave function
 \begin{align}
  \psi(x)=
  \left\{
  \begin{array}{l}
   \displaystyle
    \sqrt{\frac{2}{L}}\sin \left(-\frac{\pi}{L}x\right) \quad (-L\leq x \leq 0)\\   0 \quad (x< -L, 0 < x)
  \end{array}
  \right.\label{opt_wf}
 \end{align}
 and the minimum fluctuation $\Delta H_{A,\min}= \pi\hbar/L$, which implies
 the uncertainty relation
 \begin{align}
  \tau \Delta H_A \geq \pi\hbar\label{unc_cl}
 \end{align}
 for the ideal clock
 \footnote{We remark that this optimal wave function is the ground state of a particle confined in the infinite potential well.}
 .
 Next, we explore a Hamiltonian and a state which maximize the power output in this model.
 Since the Hamiltonian of the system is arbitrary, we set $H_S:= -C\ketbra{0} +C \ketbra{1}$ with a constant $C$. In this case, $\|H_S\|= C$.
 Let the initial state of the system be $\ket{\phi}=\ket{1}$.
 Since it is an energy eigenstate, the final energy coincides with that obtained via the unitary $U$ as mentioned in the previous paragraph.
 We choose the interaction $V_S$ so that $U=\ketbra{0}{1}+\ketbra{1}{0}$ as in \eqref{VS}.
 Then, the maximum possible work $W_{\max}=2C=2\|H_S\|$ is achieved independently of the initial state of the agent.
 According to \eqref{VS}, the size $K$ of $\mathrm{supp} V_S$ can be as small as one likes by choosing a narrowly supported function $f$.
 Thus, the interaction time $\tau$ can be arbitrarily close to the size $L$ of $\mathrm{supp} \sigma_A$.
 In this way, the minimum interaction time $\tau_{\min}=\pi\hbar/\Delta H_A$ given by \eqref{unc_cl} is achieved in arbitrary precision by setting the initial state of the agent as \eqref{opt_wf}.
 Since the maximum possible work $W_{\max}=2\|H_S\|$ is obtained independently of the initial state of the agent,
 the maximum possible power $P_{\max}$ of this model turns out to be $P_{\max}=2\|H_S\|/\tau_{\min} = 2\pi^{-1}\|H_S\|\Delta H_A/\hbar$.
 Hence, our universal upper bound $2\|H_S\|\Delta H_A /\hbar$ on the power
 is at most saturated up to the factor $\pi^{-1}$ in the ideal clock model.
 
  \paragraph{Conclusion.---}
  We have derived the universal bound (\ref{pb_f}) on the mean power produced by a quantum machine in a self-contained formulation.
  As a result of the self-contained formulation including the switch of the interaction,
  the bound shows that the energy fluctuation of the external controller is a necessary resource for producing the power.
  This is in very different circumstances from the case where we do not care about the time.
  That is, by the ideal clock model, Malabarba {\it et al.}~\cite{1367-2630-17-4-045027} showed that there is no cost to implement an arbitrary energy conserving unitary operation without caring about the time it takes.
  Hence, the possible amount of {\it the work} is correctly evaluated in such a model without consideration of the switch as we demonstrated at the beginning.
  In contrast, full consideration of the self-contained quantum machine including the switch is actually essential for the fundamental bound on {\it the power}.
  
  In addition, we have shown that this bound implies the trade-off between the power and detectability of the work when we regard the agent as the work storage.
  From this trade-off, we have derived the time scale (\ref{TS}) required for detectable work extraction in relation with the system energy scale $\|H_S\|$.
 
  We have demonstrated the ideal clock-driven quantum machine as a typical example of an autonomous quantum machine.
  In this model, we have shown that bound \eqref{pb_f} is saturated up to the factor $\pi^{-1}$.
  However, the clock is ideal because of the doubly infinite spectrum of the Hamiltonian and the perfect confinement of the initial state.
  In fact, it was shown that Condition \ref{cond1} always implies the doubly infinite energy spectrum except for the trivial case where the interaction never turns on \cite{Miyadera2016}.
  That is because Condition \ref{cond1} requires strict separation between the system and the agent before the interaction, which corresponds to the perfect confinement in the ideal clock.
  Our autonomous quantum machines are idealized in this sense.
  It is a future work to take account finite-size effects on the power bound, as Woods {\it et al.}~\cite{1607.04591} have done on the finiteness of the clock.

  Finally, we remark that our results imply that energy-time uncertainty relations have promising potential for applications in finite-time quantum thermodynamics.
  Although del Campo {\it et al.}~\cite[Supplementary Information]{Campo:2014aa} specified that quantum speed limits impose an upper bound on the output power of a quantum Otto cycle,
  physical implication of their bound is not so clear.
  Revealing a resource for producing the power, our universal bound gives evidence of the effectiveness of QSL approaches.
  Further studies are necessary for more applications.
  Structures of the system and the interaction should be taken into account for that.
  In fact, our bound do not reflect the interaction.
  Bounds in consideration of the strength of the interaction will be shown in an upcoming paper.
  
 \paragraph{Acknowledgments.---}
 KI would like to thank Masahiro Hotta and Mischa Woods for fruitful discussions and valuable comments, and acknowledges
 JSPS KAKENHI Grant Number JP16J03549. TM acknowledges JSPS KAKENHI Grant Number 15K04998.

\end{document}